\documentclass[3p,times,procedia]{elsarticle}
\usepackage{nupha_ecrc}

\volume{00}

\firstpage{1}

\journalname{Nuclear Physics A}

\runauth{J.~Berges et al.}

\jid{nupha}

\jnltitlelogo{Nuclear Physics A}

\usepackage{amssymb,amsmath}

\usepackage[figuresright]{rotating}

\def\bp{{\boldsymbol p}}
 
\def\pperp{p_{\!\perp}}

\def\bfi{\textbf{(i)}}
\def\bfii{\textbf{(ii)}}
\def\bfiii{\textbf{(iii)}}

\begin{document}

\begin{frontmatter}



\dochead{}

\title{How brightly does the Glasma shine? Photon production off-equilibrium}

\renewcommand{\thefootnote}{\fnsymbol{footnote}}

\author[ITP]{J\"{u}rgen Berges}
\author[PI]{Klaus Reygers}
\author[ITP]{Naoto Tanji\footnote{Speaker}}
\author[BNL]{Raju Venugopalan}

\address[ITP]{Institut f\"{u}r Theoretische Physik, Universit\"{a}t Heidelberg, Philosophenweg 16, 69120 Heidelberg, Germany}
\address[PI]{Physikalisches Institut, Im Neuenheimer Feld 226, 69120 Heidelberg, Germany}
\address[BNL]{Physics Department, Brookhaven National Laboratory, Bldg. 510A, Upton, NY 11973, USA}

\begin{abstract}
We present a parametric estimate of photon production at early times in heavy-ion collisions based on a consistent weak coupling thermalization scenario. We quantify the contribution of the off-equilibrium Glasma phase relative to that of a thermalized Quark-Gluon Plasma. Taking into account the constraints from charged hadron multiplicity data, 
the Glasma contribution is found to be significant especially for large values of the saturation scale.
\end{abstract}

\begin{keyword}
Photon production \sep pre-equilibrium dynamics \sep quark-gluon plasma
\end{keyword}

\end{frontmatter}

\renewcommand{\thefootnote}{\arabic{footnote}}
\setcounter{footnote}{0}

\section{Introduction} \label{sec:intro}
Photons are one of the most important probes in relativistic heavy-ion collisions as they can leave strongly-interacting matter almost unaffected and carry information that is sensitive to the different stages of the space-time evolution. 
Photon production in a thermalized Quark-Gluon Plasma (QGP) and a hadron gas has been investigated by means of  hydrodynamic and transport models~\cite{Paquet:QM17}. 
Recent photon measurements at RHIC and the LHC pose a challenge to these models in reproducing both the photon yield and elliptic flow simultaneously~\cite{Paquet:QM17,Campbell:QM17}. 
In hydrodynamic models for photon production, the contribution from the pre-equilibrium stage is not included. The understanding of pre-equilibrium photon production is of prime importance because it may open up experimental access to the early time dynamics in heavy-ion collisions. 

In the idealized high-energy limit of heavy-ion collisions, the system right after the collision is described as an over-occupied non-Abelian plasma expanding in the longitudinal direction, which is called Glasma~\cite{Lappi:2006fp}. 
Because of the over-occupation, the system is strongly interacting even though the coupling is weak. 
The real-time evolution of such a system can be computed by using classical-statistical methods. In recent classical-statistical real-time lattice simulations of the expanding Glasma~\cite{Berges:2013eia}, it has been shown that the Glasma flows to a non-thermal fixed point which corresponds to the early stage of the bottom-up thermalization scenario~\cite{Baier:2000sb}. 
In this contribution, we present a parametric estimate of photon production both in the pre-equilibrium Glasma phase and the thermal QGP phase in the bottom-up thermalization scenario in order to demonstrate the importance of the pre-equilibrium Glasma contribution~\cite{Berges:2017eom}.

\section{Estimation of the photon yields in the bottom-up thermalization scenario} \label{sec:botom-up}
In the bottom-up thermalization scenario~\cite{Baier:2000sb}, the pre-equilibrium evolution is divided into three temporal stages: \\[+2pt]
\begin{tabular}{cl}
\bfi & $Q_s^{-1} \ll \tau \ll Q_s^{-1} \alpha_s^{-3/2}$ \\
\bfii & $Q_s^{-1} \alpha_s^{-3/2} \ll \tau \ll Q_s^{-1} \alpha_s^{-5/2}$ \\
\bfiii & $Q_s^{-1} \alpha_s^{-5/2} \ll \tau \ll Q_s^{-1} \alpha_s^{-13/5}$,
\end{tabular} \\[+2pt]
where $\alpha_s$ is the QCD coupling and $Q_s$ is the saturation scale.

Stage \bfi\ is characterized by the over-occupied gluons whose typical occupation number is much larger than unity and whose typical transverse momentum is $\sim Q_s$. Due to the competition between the effects of the longitudinal expansion and the multiple two-to-two scatterings, the gluon distribution function $f_g$ shows a characteristic scaling behavior~\cite{Berges:2013eia}:
\begin{equation}
f_g (\pperp ,p_z ,\tau ) = (Q_s \tau )^{-2/3} {f}_S \left( \pperp , (Q_s \tau )^{1/3} p_z \right) \, ,
\end{equation}
where $f_S$ is a scaling function.
The overall normalization of the scaling function can be fixed by employing the results of classical Yang-Mills simulations with the color glass condensate initial condition~\cite{Krasnitz:2000gz,Lappi:2007ku}. 
To compute photon production, we need in addition to know the quark distribution $f_q$.
In Ref.~\cite{Tanji:2017suk}, it has been shown that also the quark distribution satisfies the same scaling law as the gluon distribution for typical momenta. This is because the scatterings of quarks with gluons are Bose-enhanced in a similar manner to those for gluon-gluon scattering.
We assume that $f_q$ is smaller than $f_g$ by the factor of $\alpha_s$ because the quarks obey Fermi-Dirac statistics and cannot be highly occupied. 
For photon production processes, we consider Compton scattering and the annihilation process. In the small-angle approximation, the photon production rate reads
\begin{equation}
E\frac{dN}{d^4X d^3p} = \frac{40}{9\pi^2} \alpha \alpha_s \mathcal{L}  \, f_q (\bp ) 
\int \! \frac{d^3p^\prime}{(2\pi)^3} \frac{1}{p^\prime} \left[ f_g (\bp^\prime ) +f_q (\bp^\prime ) \right] \, ,
\label{saa-rate}
\end{equation}
where $\mathcal{L}$ is the Coulomb logarithm, which corresponds to the infrared divergence of the collision processes regulated by medium mass. 
As this formula simply involves the momentum integrations of the single particle distribution functions, the photon yields $dN/dy$ can be easily evaluated. 

The photon yield from stage \bfii\ can be estimated in the same way. However in this stage, the typical occupation number of the hard gluons is less than one, and the momentum integral of the gluon distribution is dominated by soft gluons from number-changing inelastic processes. 

\begin{figure}[tb]
 \begin{tabular}{cc}
 \begin{minipage}{0.5\hsize}
  \begin{center}
   \includegraphics[clip,width=6.0cm]{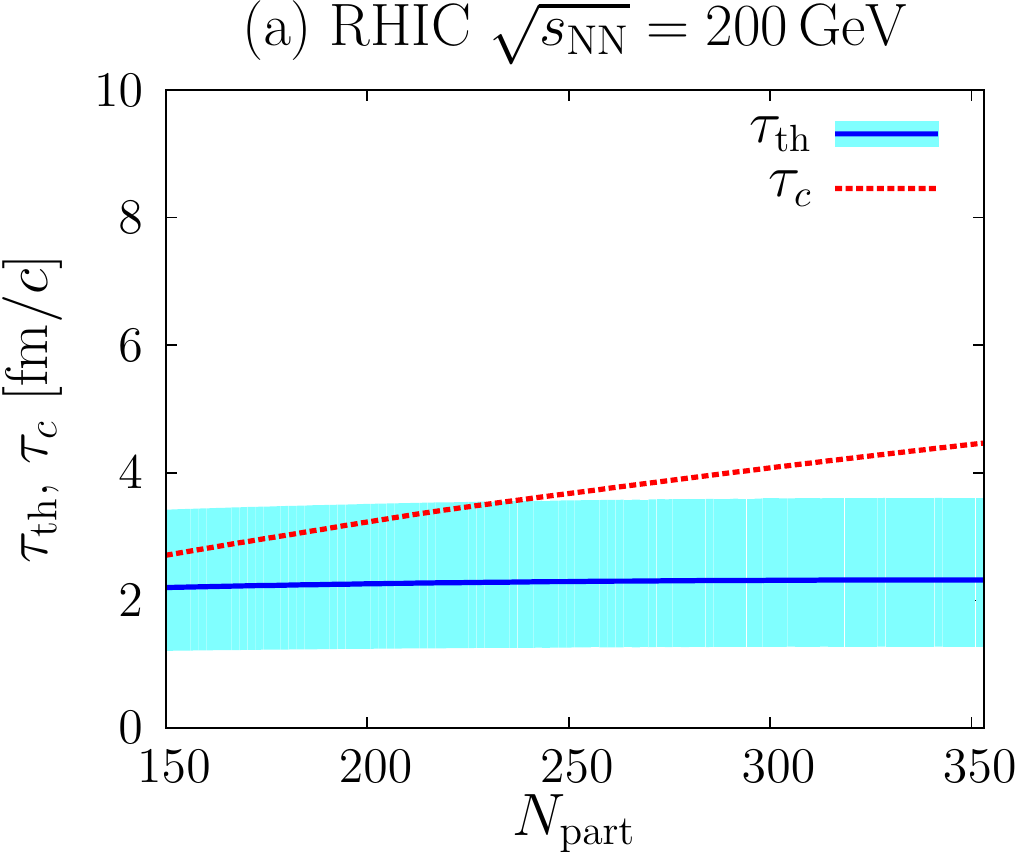} 
  \end{center}
 \end{minipage} &
 \begin{minipage}{0.5\hsize}
  \begin{center}
   \includegraphics[clip,width=5.8cm]{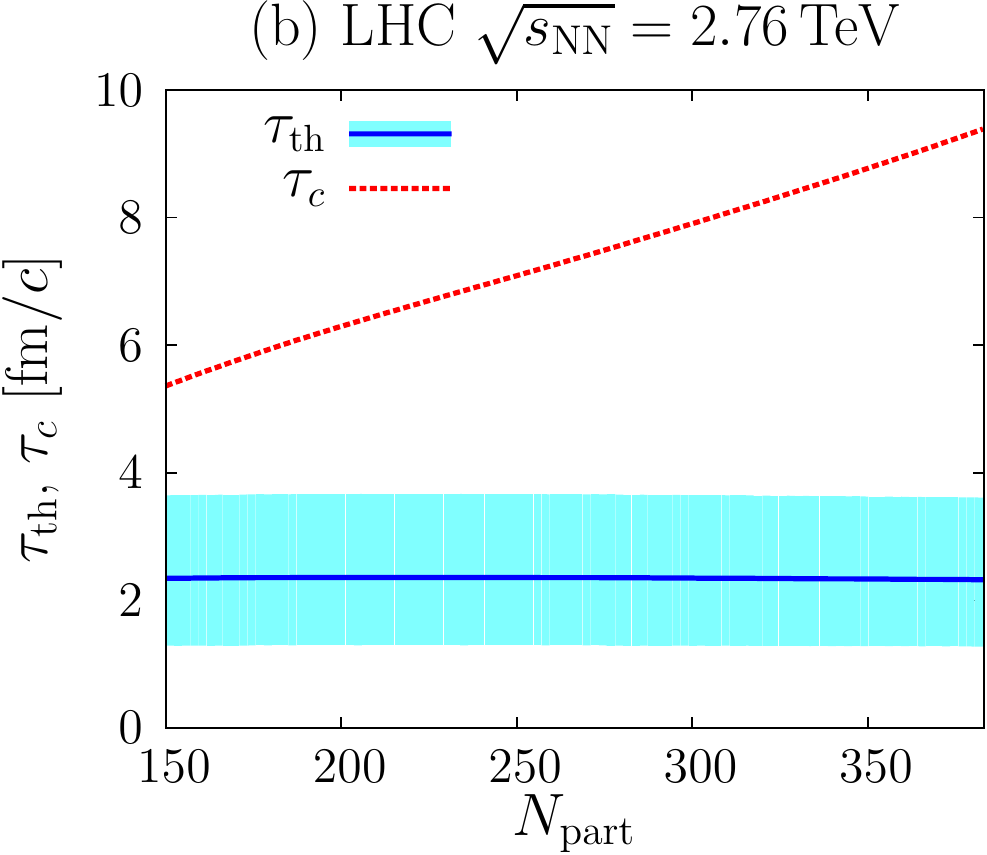} 
  \end{center}
 \end{minipage} 
 \end{tabular}
\caption{The thermalization time $\tau_\text{th}$ and the hadronization time $\tau_c$ as a function of $N_\text{part}$.
Left: RHIC $\sqrt{s_\text{NN}}=200$ GeV. Right: LHC $\sqrt{s_\text{NN}}=2.76$ TeV.}
\label{fig:tauth}
\end{figure}

In stage \bfiii, the soft gluons form a thermal bath since they have a short relaxation time compared with the typical time scale of the system. The remaining hard gluons lose their energy in this thermal bath by quenching processes and the thermal bath is heated up. 
In this study, we consider only thermal photon production from the soft thermal bath, which gives a less dominant contribution compared with those in stages \bfi\ and \bfii. 
In addition, there can be the contribution associated with the quenching of the hard quarks and gluons. We leave further discussion of this contribution to future study.

The system thermalizes by the end of stage \bfiii.
The thermalization time and the temperature at that time are parametrized, respectively, as
\begin{equation}
\tau_\text{th} = c_\text{eq} \, \alpha_s^{-13/5} Q_s^{-1} \hspace{10pt} \text{and} \hspace{10pt}
T_\text{th} = c_T c_\text{eq} \, \alpha_s^{2/5} Q_s \, ,
\end{equation}
with numerical coefficients $c_\text{eq}$ and $c_T$ denoting the uncertainty of the estimate in this scenario. 
Assuming that the system follows ideal $1+1$ dimensional expansion conserving entropy after the thermalization, we can constrain a combination of the coefficients $c_\text{eq}\, c_T^{3/4}$ by the measured charged hadron multiplicities at RHIC~\cite{Adler:2004zn} and the LHC~\cite{Aamodt:2010cz}. In Ref.~\cite{Baier:2002bt}, $c_T$ is estimated to be 0.18 up to logarithmic accuracy. To indicate the impact of the uncertainty in this quantity, we vary it by a factor of two in the range $c_T=0.1$--0.4, which is indicated by shaded bands in figures. 
The thermalization time $\tau_\text{th}$ is plotted in Fig.~\ref{fig:tauth} as a function of the number of participants $N_\text{part}$ for RHIC and LHC collision energies. It is not sensitive to $N_\text{part}$ or the collision energy. 
To evaluate the value of $\tau_\text{th}$ as a function of $N_\text{part}$, we have employed the $N_\text{part}$-dependence of $Q_s$ given by the IP-Glasma model~\cite{Schenke:2012wb}.
However, we treat the overall normalization of the values of $Q_s$ as a free parameter of our model. In Figs.~\ref{fig:tauth} and \ref{fig:photon_comp}, the normalization is chosen such that the value of $Q_s^2$ at the RHIC most central collision ($N_\text{part}=353$) is 2 GeV$^2$. 

In Fig.~\ref{fig:tauth}, we also plot the hadronization time $\tau_c$, which is defined as the time at which the temperature falls below the crossover temperature $T_c=154$ MeV. 
The photon yield in the thermal QGP phase is evaluated by integrating the thermal production rate over the expanding space-time from $\tau_\text{th}$ to $\tau_c$.

\section{Comparison of photon yields} \label{sec:results}
\begin{figure}[tb]
 \begin{tabular}{cc}
 \begin{minipage}{0.5\hsize}
  \begin{center}
   \includegraphics[clip,width=6.3cm]{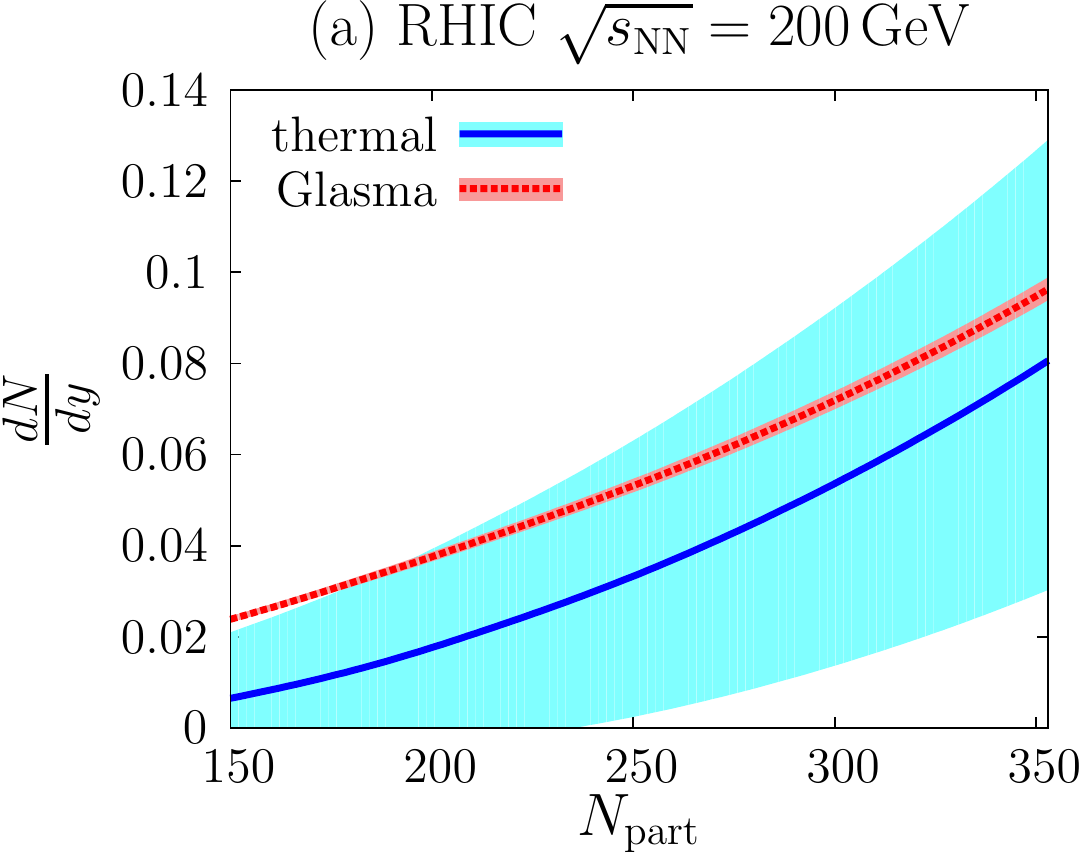} 
  \end{center}
 \end{minipage} &
 \begin{minipage}{0.5\hsize}
  \begin{center}
   \includegraphics[clip,width=6.1cm]{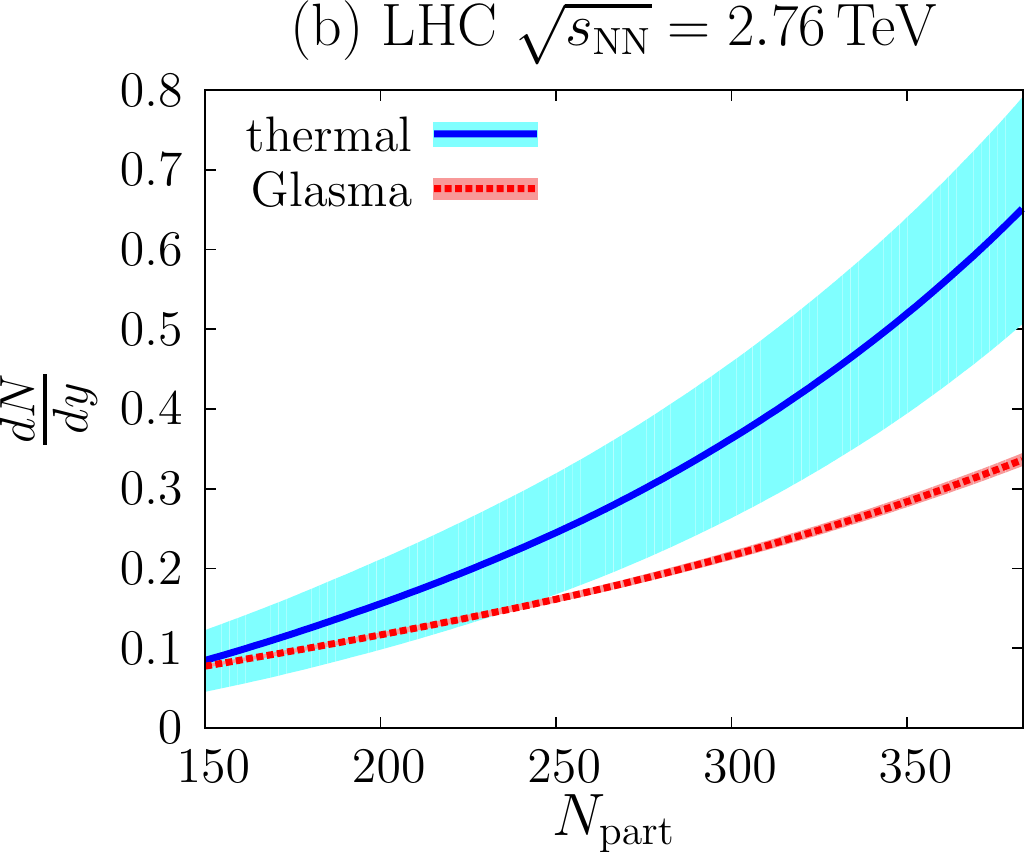} 
  \end{center}
 \end{minipage} 
 \end{tabular}
\caption{Comparison of the thermal photon yield and the pre-equilibrium Glasma photon yield in the bottom-up thermalization scenario as a function of $N_\text{part}$.
Left: RHIC $\sqrt{s_\text{NN}}=200$ GeV. Right: LHC $\sqrt{s_\text{NN}}=2.76$ TeV.}
\label{fig:photon_comp}
\end{figure}

In Fig.~\ref{fig:photon_comp}, the photon yield in the Glasma phase ($Q_s^{-1} <\tau <\tau_\text{th}$) is compared with that in the thermal QGP phase ($\tau_\text{th} <\tau <\tau_c$). 
The early-time Glasma contribution is comparable to the late-time thermal contribution even though the space-time volume is small at early times. This is because 1) the large gluon density compensates for the smallness of the space-time volume, and 2) the typical transverse momenta of quarks and gluons are larger than at later times, which enlarges the phase space volume available for photon production.
In particular, the Glasma contribution is relatively more important for lower collision energies or for less central collisions.

\begin{figure}[tb]
 \begin{tabular}{cc}
 \begin{minipage}{0.5\hsize}
  \begin{center}
   \includegraphics[clip,width=6.5cm]{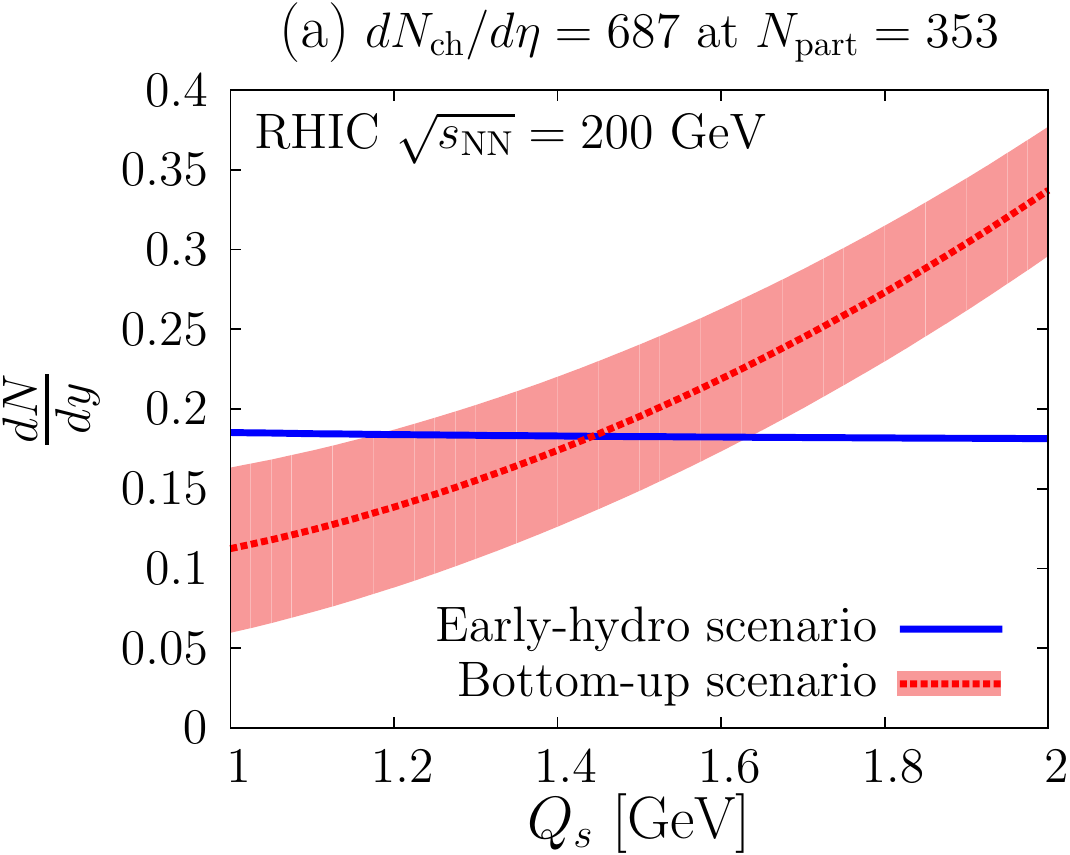} 
  \end{center}
 \end{minipage} &
 \begin{minipage}{0.5\hsize}
  \begin{center}
   \includegraphics[clip,width=6.4cm]{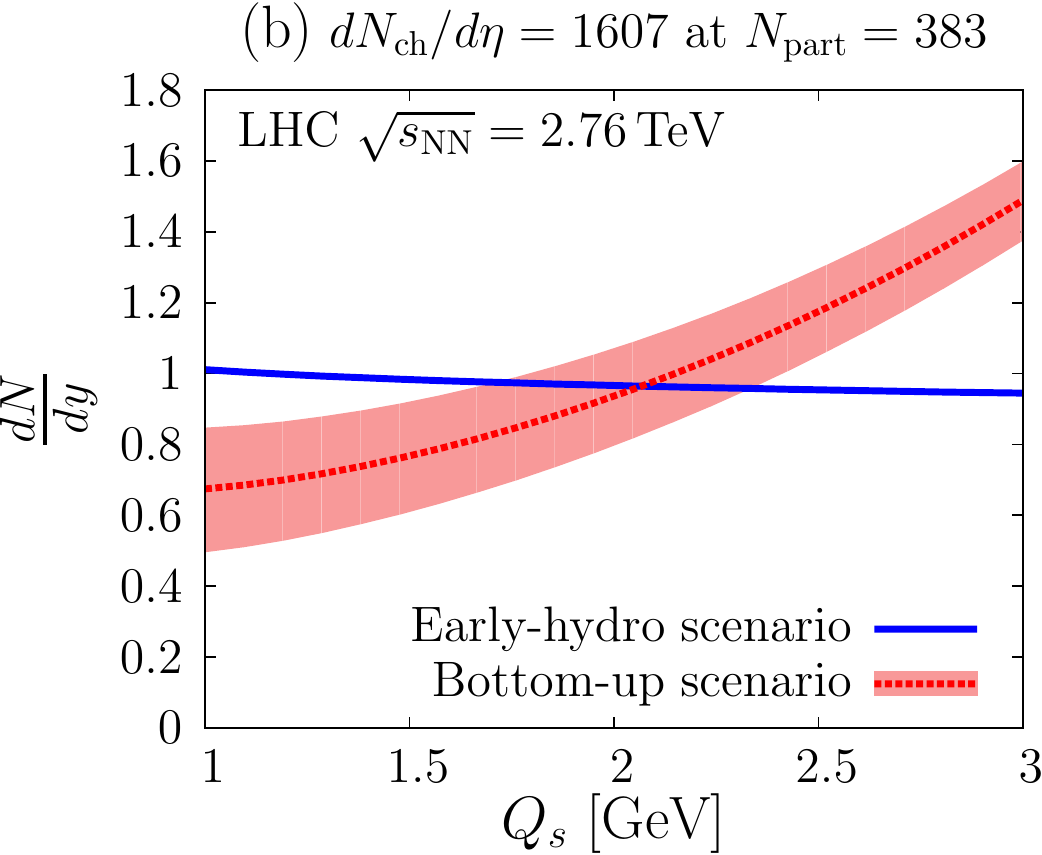} 
  \end{center}
 \end{minipage} 
 \end{tabular}
\caption{Comparison of the photon yields in the bottom-up thermalization scenario and in the hydro model extended to the early time. The results are plotted as a function of $Q_s$ for the most central collisions (centrality 0--5\%). Left: RHIC $\sqrt{s_\text{NN}}=200$ GeV. Right: LHC $\sqrt{s_\text{NN}}=2.76$ TeV.}
\label{fig:Qs-dep}
\end{figure}

In current hydro simulations for thermal photon production, the bottom-up thermalization scenario is not implemented and hydro evolution is sometimes started at early times $\sim Q_s^{-1}$. Within our simple model, the photon yield in such  early-hydro scenario can be obtained by integrating the thermal production rate from $\tau=Q_s^{-1}$ to $\tau_c$. In Fig.~\ref{fig:Qs-dep}, we compare the yield in the early-hydro scenario with that in the bottom-up thermalization scenario; the latter is given by the sum of the Glasma contribution before $\tau_\text{th}$ and the thermal contribution after $\tau_\text{th}$. The photon yields are plotted as a function of the saturation scale $Q_s$ for the most central collisions.
The yield in the early-hydro scenario is almost independent of $Q_s$ since the temperature profile is completely fixed by the hadron multiplicity data. In contrast, the yield in the bottom-up scenario has a strong dependence on $Q_s$. This is because the Glasma photon yield is approximately proportional to $Q_s^2$, which is consistent with geometrical scaling of direct photon production discussed in Ref.~\cite{Klein-Bosing:2014uaa}.
If the value of $Q_s$ is larger than $\sim 1.5$ GeV for RHIC and $\sim 2$ GeV for the LHC, the bottom-up thermalization  produces more photons than the hydro model extended to early times.

\section{Conclusions} \label{sec:conclusions}
In this work, we discussed parametric estimates of photon yields in the different stages of the bottom-up thermalization scenario and demonstrated the importance of the pre-equilibrium Glasma contribution for photon production.
These results motivate more rigorous ab-initio calculations of photon production at early times in heavy-ion collisions which employ the kinetic theory or classical-statistical simulations. 
Those computations may reduce the uncertainties we identified in our estimates and enable us to access the momentum spectrum of direct photons, which is crucial in order to address the photon $v_2$ puzzle.

\bigskip 
This work is part of and supported by the DFG Collaborative Research Centre ``SFB~1225~(ISOQUANT)". 
R.~V.~is supported under DOE Contract No.~DE-S{C001}2704.






\end{document}